\documentclass[12pt]{article}
\begin{document}
\hfill {\bf DO-TH 14/29} \\
\vspace{2.5cm}

\begin{center}
{\bf\Large{On the Proton charge extensions}}\\

\bigskip
\bigskip
{\bf{M.~Gl\"uck}}\\

\bigskip
{\it{Institut f\"ur Physik, Technische Universit\"at Dortmund,}}\\
{\it{D-44221 Dortmund}}
\end{center}

\vspace{2.5cm}

\begin{center}{\bf{Abstract}}\end{center}

\medskip

\noindent It is shown that the recent determination of the various proton charge extensions is compatible with Standard Model expectations.

\newpage

A proton charge extension is defined according to its impact on some specific measurement.
The impact of the parameter $r_p$ in
\begin{equation}
F_1^p(\vec q\,^2) = 1 -\frac{\vec q\,^2}{6} r^2_p
\end{equation}
on the energy level $E_n$ of the hydrogen atom is commonly given  by [1]
\begin{equation}
\Delta E_n = \frac{2}{3}\, \pi\alpha\,  |\psi_n(0)|^2 \, r_p^2
\end{equation}
It turns out, however, that $r_{p,\mu}< r_{p,e}$, which is considered [2] as a `Proton Radius Puzzle'. 

Considering the different extensions of the muon and the electron wave functions in the corresponding hydrogen atoms, it is
clear that the parametrization of $F_1^p(\vec q\,^2)$ should depend on {\underline{two}} free parameters, e.g. 
\begin{equation}
F_1^p(\vec q\,^2) = 1 -\frac{\vec q\,^2}{6} \, r^2_p + \frac{(\vec q\,^2)^2}{6} \, \tilde{r}_p^4 \, ,
\end{equation}
implying that 
\begin{equation}
\Delta E_{n,\ell} = \frac{2}{3}\, \pi\alpha |\psi_{n,\ell}(0)|^2 \,\,  r_{p,\ell}^2
\end{equation}
where $\ell=e,\mu$ and where
\begin{equation}
r_{p,\ell}^2 = r_p^2 -\langle\vec q\, ^2\rangle_{n,\ell}\,  \tilde{r}_p^4\, ,
\end{equation}
with
\begin{equation}
\langle\vec q\, ^2\rangle_{n,\ell} = |\psi_{n,\ell}(0)|^{-2} \int \frac{d^3q}{(2\pi)^3}\, \vec q\, ^2 \, 
                   \int d^3re^{i\,\vec q \cdot \vec r} |\psi_{n,\ell}(\vec r)|^2 \, .
\end{equation}
\\
Thus $\langle\vec q\, ^2\rangle_{n,\mu}  >  \langle\vec q\, ^2\rangle_{n,e}$, accounts for the observed [2]
$r_{p,\mu}< r_{p,e}$,  in full agreement with Standard Model expectations.

\vspace{1.25cm}
\noindent{\bf Acknowledgement}

\noindent I am indebted to Prof.~E.~Reya for valuable information and remarks.

\vspace{2.0cm}

\noindent{\bf References}
\begin{enumerate}
\item{R.~Karplus, A.~Klein, and J.~Schwinger, Phys.~Rev. {\bf 86} (1952) 1183.}
\item{R.~Pohl et al., Ann.~Rev.~Nucl.~Part.~Sci. {\bf 63} (2013) 175, and references therein.} 
\end{enumerate}

\end{document}